\documentstyle[array,multicol,aps,rmp,psfig]{revtex}

\begin{document}

\title{\Large \bf Binocular  disparity  can   explain  the 
orientation  of  ocular  dominance   stripes  in  primate  V1}
\author{Dmitri B. Chklovskii}
\address{Sloan Center for Theoretical Neurobiology, The Salk Institute, La Jolla, CA 92037}

\date{\today}
\draft
\tighten
\preprint{Submitted to {\it Vision Research}}
\maketitle

\begin{abstract}
In the primate primary visual area (V1), the ocular dominance pattern
consists of alternating monocular stripes. Stripe orientation follows
systematic trends preserved across several species. I propose that
these trends result from minimizing the length of intra-cortical
wiring needed to recombine information from the two eyes in order to
achieve the perception of depth. I argue that the stripe orientation
at any point of V1 should follow the direction of binocular disparity
in the corresponding point of the visual field. The optimal pattern
of stripes determined from this argument agrees with the ocular
dominance pattern of macaque and Cebus monkeys. This theory predicts
that for any point in the visual field the limits of depth perception
are greatest in the direction along the ocular dominance stripes at
that point.
\end{abstract}

\begin{multicols}{2}

\section{INTRODUCTION}

\begin{figure}
\centerline{
\psfig{file=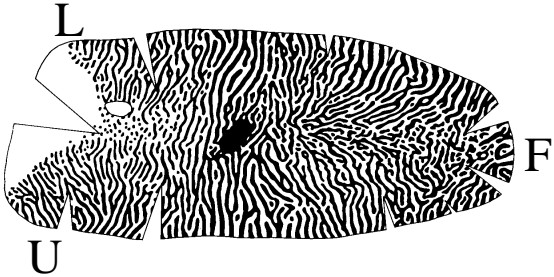,width=2.9in}
}
\vspace{0.1in} 
\setlength{\columnwidth}{3.4in}
\caption{The complete reconstruction of the ocular dominance
pattern in macaque V1 by LeVay et al. Black/white stripes are composed
of neurons which are dominated by the left/right eye. The retinotopic
map is organized as follows.  Most of the perimeter is the V1-V2 border
which corresponds to the vertical meridian the upper end of which is 
marked by $U$ and the lower by $L$. $F$ designates the
representation of the fovea.  The horizontal meridian is represented
along the long axis with the tilted black oval corresponding to the
blind spot.
\label{fig:LVa}
}
\vspace{-0.1in}
\end{figure}

The perception of depth in primates relies on recombining information
coming from both eyes. This is accomplished by a retinotopic mapping
\cite{DanWhitt},\cite{Tootell} of the two retinal images onto the
primary visual area V1. In many primates, neurons dominated by each
eye are segregated into the system of alternating stripes known as the
ocular dominance pattern \cite{HubWies68},\cite{WiesHub74}. A complete
reconstruction of the pattern in a macaque V1 is shown in
Fig.\ref{fig:LVa}.\cite{LeVay}

\begin{figure}
\centerline{
\psfig{file=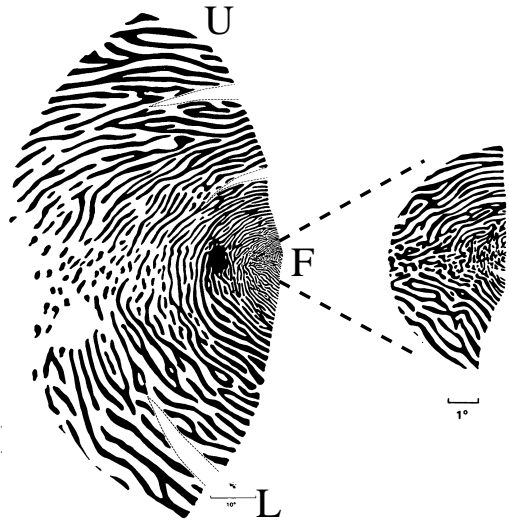,width=2.9in}
}
\vspace{0in} 
\setlength{\columnwidth}{3.4in}
\caption{The complete ocular dominance pattern from Fig.\ref{fig:LVa}
transformed back into visual field coordinates by LeVay et al. The
right boundary represents the vertical meridian with the fovea (F) in
the middle. Notice two major trends in the orientation of the stripes.
In the parafoveal region stripes tend to run horizontally. Farther
from the fovea stripes follow concentric circles. The blow-up shows
parafoveal region.
\label{fig:LVb}
}
\vspace{-0.1in}
\end{figure}

Although this pattern may seem random, the orientation of the ocular
dominance stripes on the cortical surface follows systematic trends
found in other macaques \cite{HortonVary} and in Cebus
monkeys \cite{Rosa}. These trends are easiest to see when the ocular
dominance pattern is transformed back into visual field coordinates by
dividing all cortical distances by the local magnification
factor. (The magnification factor \cite{DanWhitt} gives the distance in
millimeters on the cortex which corresponds to a $1\deg$ separation on
the retina.)  The transformed pattern shown in Fig.\ref{fig:LVb} (as
obtained by LeVay {\em et al.} (1985) following Hubel and
Freeman (1977)) reveals two major trends: in the parafoveal
region stripes tend to run horizontally, while farther from the fovea
stripes follow roughly concentric circles. These trends in the
orientation of the stripes call for explanation.

\vspace{-0.1cm}

Many theorists have successfully modeled the development of ocular
dominance columns \cite{Erwin},\cite{SwinRev},\cite{Wisk}, and some
have addressed the trends in stripe orientation. It was suggested
originally by LeVay {\em et al.} (1985) and later investigated by
others \cite{Jones},\cite{GoodWill},\cite{Bauer} that the mapping from
the two almost circular LGN layers to the more elongated
representation in V1 requires least stretching (or anisotropy of the
magnification factor within ocular dominance stripes) if the stripes
run perpendicular to the long axis of V1. However, this theory does
not explain the different orientation of ocular dominance stripes in
the parafoveal region.\cite{GoodOrient} Moreover, it is unlikely that
the shape of V1 dictates its internal organization; more probably, its
internal organization will dictate its shape. Goodhill {\em et
al.} (1997) have instead proposed that the global pattern of
ocular dominance stripes arises from anisotropic and spatially
non-uniform correlations in the neural input from the retinae. This
seems like a plausible developmental mechanism, although anisotropic
correlations have yet to be demonstrated experimentally.

Rather than modeling development, I have taken a different approach to
explain the orientation of ocular dominance stripes. I propose that
the stripe orientations follow from V1's role in depth perception
according to the principle of wiring economy. In other words, I focus
on understanding {\it why} the stripes are arranged as they are,
rather than {\it how} they become so arranged.

The wiring economy principle amounts to the 
following \cite{Cajal},\cite{Cowey},\cite{Mitch1},\cite{Chern},
\cite{Young},\cite{Cat}: because of limitations on head size, there is
pressure to keep the volume of the cortex to a minimum. This implies
that wiring, i.e. axons and dendrites, should be as short as possible,
while maintaining function. In general, the function of a given
cortical circuit specifies the connections between neurons.  Therefore
the problem presented by the wiring economy argument is to find, for a
given set of connections, the spatial layout of the neurons that
minimizes wiring length. This problem can be extremely difficult
because of the large number of neurons and even larger number of
interconnections within a cortical region. However, the columnar
organization of the cortex \cite{Mou} allows me to consider the layout
of cortical columns (each consisting of $\approx 10^4$ neurons) rather
than individual neurons, reducing the problem to two dimensions.

The wiring economy principle has been used to explain retinotopic
map \cite{Cowey},\cite{Kaas} and ocular dominance
stripes \cite{Mitch1},\cite{OD} in V1. In the first case, the
construction of receptive fields requires connections between neurons
representing neighboring points in the visual field. Topographic
mapping of the visual field minimizes the length of these
connections. In the second case, each cortical neuron connects more
often to cortical neurons dominated by the same eye than to neurons
dominated by the opposing eye. Thus the segregation of neurons into
alternating monocular stripes minimizes the total length of
intra-cortical connections.

Here I use the principle of wiring economy to find the optimal
orientation of the ocular dominance stripes (given they exist) from
the function of V1 in processing binocular disparity. Disparity arises
when an object closer or farther than the point of fixation forms
images on the two retinas that are in different positions relative to
the fovea.  Because the cortex is retinotopically mapped, the left and
the right eye representations of this object in the cortex will fall
some distance away from each other, as determined by the magnitude and
direction of the retinal separation. Recombining these representations
requires extensive wiring between cortical columns.

In the next Section, I show that the left and right eye
representations in V1 will be farther apart if the ocular dominance
stripes run perpendicular, rather than parallel, to the direction of
retinal separation.  Therefore I argue that in order to minimize
wirelength, the orientation of the stripes should correspond to the
direction of binocular disparity. However, for any given point on the
cortex the disparity in the corresponding point of the visual field
depends on the viewing conditions, that is, direction of gaze and
distance to the object.  In Section \ref{sec:map}, I calculate a distribution of
disparities averaged over viewing conditions. I obtain a map of
typical disparities in the visual field which then determines the
optimal orientation of the ocular dominance stripes.

The orientation of stripes predicted by this theory agrees with actual
patterns obtained from macaque and Cebus monkeys, see Section \ref{sec:dis}. 
My results show that
the two major trends of stripe orientation result from two main
contributions to disparity. In the parafoveal region, binocular
disparity is due mainly to the horizontal displacement of the eyes,
consistent with the horizontal stripes in the (transformed) ocular
dominance pattern. Farther from the fovea, the pattern consists of
isoeccentric lines. These are explained by binocular disparity due to
unequal rotation of the eyeballs around the gaze line (cyclotorsion).

\section{Disparity  direction  determines  orientation  of  ocular 
 dominance  stripes}

Because the left and the right eye pathways do not converge before V1,
the existence of binocular neurons in V1 \cite{HW70} suggests that the
information from both eyes is recombined there. Moreover, many
binocular neurons in V1 are disparity-tuned. \cite{BBP},\cite{PF} This
requires intra-cortical wiring which connects cortical columns
containing left/right eye representations of an object. To minimize
the wirelength, the distance between these columns should be as small
as possible.

\begin{figure}
\centerline{
\psfig{file=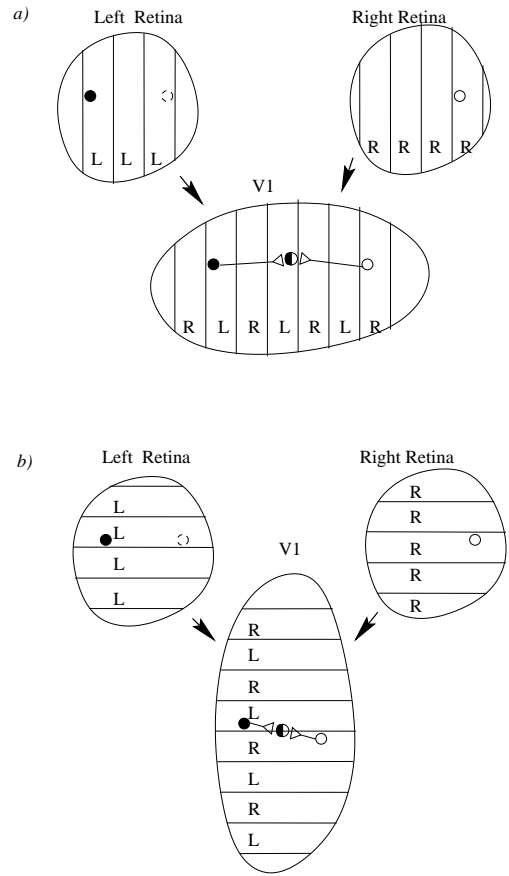,width=2.6in}
}
\vspace{0.1in} 
\setlength{\columnwidth}{3.4in}
\caption{Alignment of ocular dominance stripes with the disparity
direction in the cortex minimizes wirelength. Filled and empty circles
designate left and right retinal images of the same object.  Dashed
circle is the point of the left retina corresponding to the right-eye
image. Separation between the filled and the dashed circle is the
retinal disparity. Shown are the two limiting cases of the
retino-cortical mapping with ocular dominance stripes perpendicular
(a) or parallel (b) to the disparity direction. The two images are
recombined in V1 (for example by projecting onto a binocular cell,
half-filled circle). Because of the double-coverage perpendicular to
the ocular dominance stripes, alignment of ocular dominance stripes
with the disparity direction (b) places the right/left representations
of the same object closes than (a).
\label{fig:wire}
}
\vspace{0.1in} 
\end{figure}

For a given magnitude of binocular disparity, the distance between the
columns containing left/right representations of an object depends on
the orientation of ocular dominance stripes relative to the separation
of the columns. To see this consider two alternative arrangements:
ocular dominance stripes oriented perpendicular, Fig.\ref{fig:wire}a,
or parallel, Fig.\ref{fig:wire}b, to the separation of the columns.
One can think of V1 as being composed of interleaved stripes cut from
the two topographic maps belonging to the two eyes.\cite{BlaFitz} If
one were to move across the stripes the representation of every point
in the visual field is encoutered twice: once in a right-eye column,
once in a left-eye column. Therefore the separation between the two
columns containing left/right eye representations of the object is
twice as big if the stripes are perpendicular compared to parallel to
the separation between the columns. Thus for a given magnitude of
disparity in the visual field the length of inter-eye connections is
minimized if the ocular dominance stripes run in the direction
corresponding to the disparity direction.

Several assumptions were made in this argument. First, I assumed the
absence of stretching within the stripes which could change the
distances across vs. along the stripes. To see whether this is a valid
assumption I restate the argument in terms of the cortical
magnification factor which has been measured experimentally. The
separation between the two columns in V1 is given by the retinal
disparity times the magnification factor. If the magnification factor
across the stripes is greater than that along the stripes, the
separation between the columns and hence the length of inter-eye
connections is minimized by aligning the stripes with the disparity
direction. Experiments show that the cortical magnification factor
along the stripes is, indeed, about 1.5 times smaller than that across
the stripes.\cite{Blasdel} Therefore, even though some stretching
exists, my argument remains valid.

Second, I did not include wirelength of intra-eye connections in the
cortex. These connections are responsible for monocular functions of V1 such as
processing of contour orientation and color. The reason for
neglecting these connections is their isotropy, that is they do not
depend on the direction in the visual field. Therefore, orientation of
the ocular dominance stripes should not affect the length of intra-eye
connection.

Third, I neglected a possible specificity of inter-eye connections
in respect to monocular functions of V1 such as orientational selectivity.
Binocular neurons are likely to receive information from the neurons
with the same preferred orientation. Moreover this preferred
orientation should correlate with the disparity direction. However,
this effect should not affect my argument because it averages
out. Once all the possible orientations are included, the combined
connections should be non-specific because different orientations are
approximately equally represented.

Fourth, I assumed retinotopic mapping in V1. Although there is scatter
in the receptive field location in a given cortical column, the
magnitude of the scatter does not exceed the period of the ocular
dominance pattern.\cite{HubWies74} Because I rely on retinotopy on the
scales of several stripe widths (Fig.\ref{fig:wire}) the argument
remains valid.

Thus I showed that the orientation of ocular dominance 
stripes should follow the direction of disparity for the corresponding 
point of the visual field. To determine the optimal pattern of ocular
dominance stripes I need to find disparity for all points in the 
visual field. Thus I need to calculate a binocular disparity map.

\section{Calculation  of  the  typical  disparity  map}
\label{sec:map}

In the previous Section, I argued that the ocular dominance pattern
should follow the map of binocular disparity in order to minimize the
length of intra-cortical wiring.  However the direction and magnitude
of the disparity for a given point in the visual field depends on the
viewing conditions such as the distance to the object and the
direction of gaze. Therefore I need to average disparity over these
variables. The typical direction of disparity should determine the
optimal orientation of ocular dominance stripes.

In order to find the disparity map I consider the origins of disparity
in some detail. This is done in several steps by first considering a
primate with the gaze direction fixed at straight ahead and
then gradually adding degrees of freedom available to the eyeballs.

\begin{figure}
\centerline{
\psfig{file=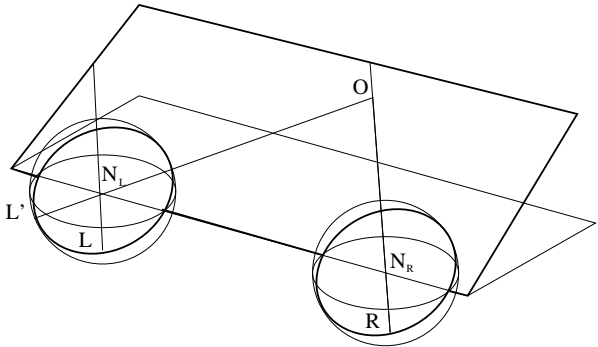,width=3.1in}
}
\vspace{0.1in} 
\setlength{\columnwidth}{3.4in}
\caption{Epipolar lines (thick circles) determine the direction of
disparity for fixation at optical infinity. They are formed by the
intersection of the retinae and an epipolar (visual) plane (thick
rectangle), which passes through fixation point and the centers of the
two eyeballs.
\label{fig:geom}
}
\vspace{0.1in} 
\end{figure}

Consider two eyes fixating at optical infinity. Then, by definition,
the images of the fixation point fall on the foveae of the two
eyes. Moreover, all objects at infinity are imaged on the retinal
locations which are the same distance and direction from the fovea in
both eyes. Such locations send afferents to adjacent cortical columns
and are called corresponding. (In reality, images of infinite objects
may not fall on exactly corresponding points. For example, there is a
$2\deg$ tilt of the vertical
meridians.\cite{Volkmann},\cite{Helmholtz} In this paper I neglect
these deviations because they do not alter the results qualitatively.)
Physiologically, stimulation of corresponding points results in a
single perception of the object. Objects at finite distance away,
however, are imaged at different retinal locations relative to the
fovea, called non-corresponding.  Binocular disparity is defined as a
displacement of the left-eye image from the location on the left
retina corresponding to the right-eye image of the same point object.
Physiologically, finite-distance objects may still appear single due
to sensory fusion if the disparity falls within a range called Panum's
fusional area. Otherwise the doubling of the perception or diplopia is
experienced.

When the eyes fixate at optical infinity all the infinitely removed
objects appear with zero disparity. To find disparity of other objects 
I use a geometrical construction illustrated in
Fig.\ref{fig:geom}. I fix the direction of gaze at straight ahead.
The image of point $O$ in the right eye falls on the retinal point $R$
which belongs to the line passing through $O$ and the nodal point of
the right eye $N_R$. The image of point $O$ in the left eye falls on
the retinal point $L'$. A line passing through the left-eye nodal
point $N_L$ and parallel to $OR$ intersects the retina at the point
$L$, a point corresponding to $R$. The arc $LL'$ is the binocular
disparity of point $O$. This arc belongs both to the retina and to a
plane that passes through point $O$ and the nodal points of the two
eyes, known as an epipolar plane. The common of the epipolar plane and
the retina is called an epipolar line. Therefore the direction of
disparity $LL'$ is along the epipolar line while its magnitude depends
on the distance to point $O$. If point $O$ had a different elevation
its disparity direction will be aligned with another epipolar line
formed by the intersection of another epipolar plane and the
retina. Therefore possible directions of disparity in the visual field
are along epipolar lines formed by great circles passing through the
interocular line. I call this disparity translational because it results
from the horizontal displacement of the eyes.

Now I allow eyes to change the direction of gaze in the horizontal
plane, while assuming that the center of rotation of an eyeball
coincides with its nodal point. Then point $O$ is projected onto the
same locations $L'$ and $R$ in head-centered coordinates. Point $L$
remains corresponding to $R$. However the eyes, and hence the retinae
rotate under those points. Therefore, the direction of disparity in
the retinal coordinates changes depending on the gaze direction. I
calculate the frequency distribution of disparity directions by
averaging over a uniform distribution of gaze directions within $\pm
30\deg$ of straight ahead. 

\begin{figure}
\centerline{
\psfig{file=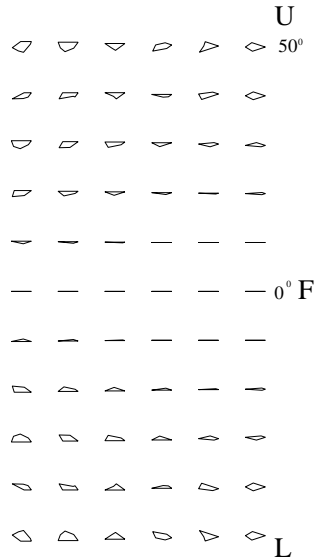,width=1.6in}
}
\vspace{0.1in} 
\setlength{\columnwidth}{3.4in}
\caption{Weighted disparity direction map for the left visual
hemifield calculated from Listing's law for a distribution of
directions of gaze with an azimuth and within $30\deg$ from the
primary position. The grid of polar plots is arranged in the
azimuth-elevation plane of the retina. Each plot represents the
distribution of disparity directions for a corresponding point on the
reina.
\label{fig:listb}
}
\vspace{0.1in} 
\end{figure}

The result is shown in Fig.\ref{fig:listb}
as a grid of polar plots each of which corresponds to a particular
point in the retinal coordinates.  Each polar plot shows the frequency
of different directions of disparity for a given point on the
retina. The distribution of disparity exhibits strong anisotropy and
the dominant disparity direction can be easily determined for all the
retinal locations.  If the only movements allowed to the eye were
rotations around vertical axis then this would be a complete disparity
map. Optimal ocular dominance pattern would be determined by
transforming this map into cortical coordinates by using the
magnification factor. This map agrees with the parafoveal stripe
orientation in macaque but disagrees with it farther from the fovea.

Inclusion of different gaze elevations eliminates this disagreement.
Naively, one may expect that the disparity map remains intact because
the interocular line is the axis of rotational symmetry. However,
vertical eye movements are accompanied by
cyclotorsion,\cite{Nakayama},\cite{Enright} or rotation of the
eyeballs around the direction of gaze. The amplitude of cyclotorsion
depends on the direction of gaze as specified by Listing's
law. (Listing's law states that to determine the amplitude of
cyclotorsion for an arbitrary direction of gaze one has to rotate the
eye into that direction from the primary position around an axis which
lies in a (Listing) plane.) According to the recent
measurements \cite{Bruno} the amplitude of cyclotorsion is often
unequal in the two eyes.  Thus, although points $L'$ and $R$ remain
fixed in the head-centered coordinates, point $L$ corresponding to $R$
rotates around the direction of gaze direction. This causes a
cyclotorsional contribution to disparity which is oriented along
concentric circles around the fovea.

The full binocular disparity includes cyclotorsional and translational
contributions. Because translational contribution depends on the
distance to an object while cyclotorsional does not, the direction of
disparity depends on the distance to an object. Therefore, finding the
typical disparity direction requires specifying a range of distances
to objects that are perceived stereoscopically.

\begin{figure}
\centerline{
\psfig{file=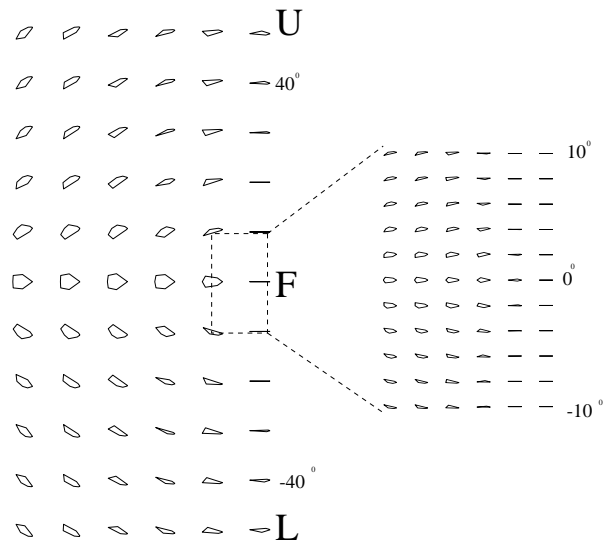,width=3.1in}
}
\vspace{0.1in} 
\setlength{\columnwidth}{3.4in}
\caption{Weighted disparity map for a distribution of gaze
directions within $30\deg$ of the primary position. The grid of polar
plots showing the distribution of disparity directions is in
 azimuth-elevation visual field coordinates. Both cyclotorsional and
translational components of disparity are included in the calculation.
Notice that the dominant disparity directions are similar to the
pattern of ocular dominance in Fig.\ref{fig:LVb}. The blow-up shows 
parafoveal region.
\label{fig:DCa}
}
\vspace{-0.1in} 
\end{figure}

To define this range of distances I use the following argument.
Because the cost of connections grows with their length, the pattern
of connections in the cortex should not be far from isotropic. For a
given location on the cortex, the circle of intra-cortical connections
corresponds to an ellipse in the visual field because of the
anisotropic magnification factor. The short axis of this ellipse $a$
is perpendicular to the direction corresponding to the stripes while
the long axis $b$ is parallel.  Information is recombined from all
objects with disparity less than $a$ and is not recombined from any
object with disparity greater than $b$.  Whether information is
recombined from objects with disparity in the interval $[a;b]$ depends
on the orientation of ocular dominance stripes.  In order to choose
the optimal direction I need to calculate the frequencies of disparity
directions for the object locations whose magnitude of disparity falls
in the interval $[a;b]$. I choose the values of $a$ and $b$ as explained
in Methods. Although, results of the calculation depend on the choice
of parameters, they do not change qualitatively (see Discussion).
The typical disparity map is shown in Fig.\ref{fig:DCa}.

\section{Discussion}
\label{sec:dis}

Orienting ocular dominance stripes in the direction locally
corresponding to the typical disparity optimizes the length of
intra-cortical wiring needed for the perception of depth. Therefore, the
wiring economy principle predicts that the ocular dominance pattern
follows the map of typical disparities. This prediction agrees with
the data as can be seen by comparing the map of typical disparities in
Fig.\ref{fig:DCa} with the data of LeVay {\em et al.} (1985),
Fig.\ref{fig:LVb}. The map of typical disparities reproduces
correctly the two major trends in the data: in the parafoveal region
stripes tend to run horizontally, farther from the fovea, the pattern
consists of isoeccentric stripes.

\begin{figure}
\centerline{
\psfig{file=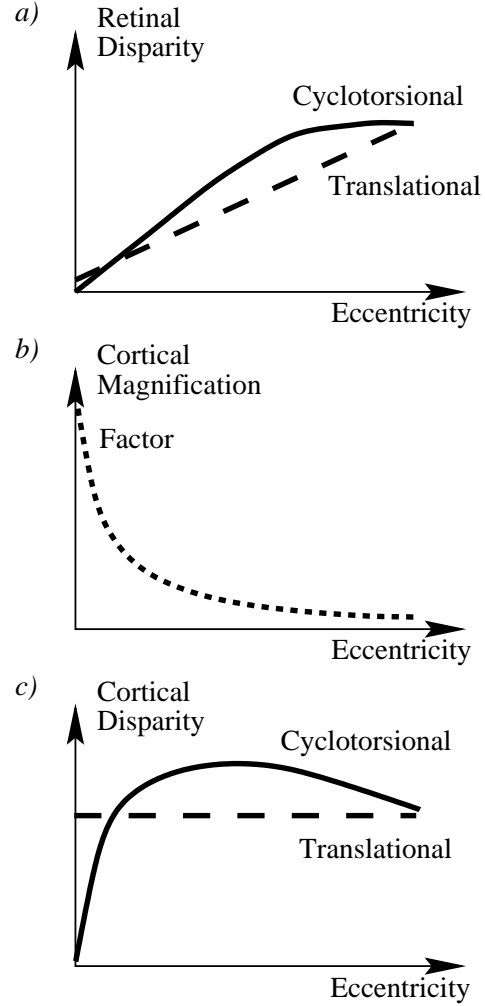,width=2.5in}
}
\vspace{0.1in} 
\setlength{\columnwidth}{3.4in}
\caption{Relative importance of the two contributions to disparity
depend on the eccentricity. a)Magnitude of translational and
cyclotorsional disparity components in retinal coordinates as a
function of eccentricity along the horizontal meridian. Notice that
the cyclotorsional disparity goes to zero linearly with eccentricity
while the translational disparity remains finite in the
parafovea. b)Cortical magnification factor as a function of
eccentricity. c)Translational and cyclotorsional disparity components
in cortical coordinates as a function of eccentricity along the
horizontal meridian.
\label{fig:hor}
}
\vspace{-0.1in} 
\end{figure}

These trends result from the two major components of disparity:
translational, due to the horizontal displacement of the eyes, and
cyclotorsional, due to unequal rotation of the eyeballs around the
gaze line (cyclotorsion). The relative magnitude of the two components
depends on the distance to the fovea, Fig.\ref{fig:hor}.  In the
parafoveal region, cyclotorsional disparity goes to zero linearly with
eccentricity because rotational displacement is proportional to the
radial distance from the axis of rotation passing through the fovea.
At the same time, translational disparity remains finite for objects
closer or farther than the point of fixation. Therefore, translational
disparity dominates in the parafoveal region. Since the translational
disparity is mostly horizontal, this explains the horizontal trend in
the stripe orientation in parafoveal region. Farther from the fovea,
cyclotorsional component of disparity may (or may not) become dominant
depending on several parameters: the amplitude of cyclotorsion,
the frequency of different viewing conditions, and the typical limit of
depth perception.  Along the horizontal meridian, cyclotorsional
disparity is vertical, while translational is horizontal. Hence, the
direction of the ocular dominance stripes must switch at the point
where cyclotorsional component of disparity takes over the
translational.  This switch is evident in the macaque data,
Fig.\ref{fig:LVb} at about $8\deg$ eccentricity.

These trends in the orientation of ocular dominance stripes are not
qualitatively affected by assumptions made in the calculation.  For
example, I assumed a uniform distribution of the gaze directions.  Any
reasonable bell-shaped distribution should lead to the same two trends
in the ocular dominance pattern.  Although I considered fixation at
infinity my results are qualitatively correct for fixation at nearby
objects. A simple geometrical argument shows that the translational
disparity has a greater vertical component for close distances. Also,
cyclotorsional disparity is greater because cyclotorsion increases
with the vergence angle.\cite{Mok},\cite{Van},\cite{Minken}
Qualitatively, the ocular dominance pattern displays the same two
trends. However the switch in the stripe orientation may occur at
different eccentricity.

Generality of the trends in the stripe orientation 
confirms their functional significance. Ocular dominance
patterns imaged in several macaques \cite{HortonVary} show the two
trends in the stripe orientation. Rosa {\em et al.} (1992) transformed
into visual field cooridinates a complete ocular dominance pattern of
{\it Cebus} monkeys. They found that the pattern was qualitatively
similar to macaque with the stripe orientation switching at $\approx
6\deg$ and the first trend sometimes lacking. Preliminary
data \cite{Horpriv} on the ocular dominance pattern in humans is hard
to analyze because precise topography in V1 is not known. Although the
two trends are present, there is a significant interpersonal
variability, possibly indicating varying significance of the two
contributions to disparity from person to person.

Although, this theory reproduces the two major trends in the data, there
is an unexplained trend in the macaque \cite{HortonVary} and {\it
Cebus} monkey \cite{Rosa} data. In the foveal region, less than $1\deg$
eccentricity, the orientation of stripes differs from the typically
horizontal disparity there.  I speculate that this may be due to
fixation point being projected away from the V1-V2 border. To verify
this suggestion a combined topographic and ocular dominance mapping is
needed.

This theory relates ocular dominance pattern to the function of V1
allowing me to make several predictions. The location of the switch in
the orientation of the ocular dominance stripes along the horizontal
meridian should depend on the following parameters. Greater amplitude
of cyclotorsion (or a greater frequency of gaze directions requiring
cyclotorsion) increases cyclotorsional disparity and pushes the
location of the switch in the stripe orientation towards the fovea.  A
greater extent of the Panum's fusional area, $b$, achieves fusion for
more objects with largely translational disparity. This should
increase the eccentricity of the switch. Schwartz (1980)
suggested that the size of the Panum's fusional area and the width of
the ocular dominance stripes are correlated between different
species. If this is correct my theory implies that the species with
greater width of ocular dominance stripes in the visual field should
have the switch at higher eccentricity.

According to the theory, the functional significance of the
orientation of ocular dominance stripes is in accommodating the
typical disparity direction. Then processing of binocular disparity
for any point of the visual field should be more efficient in the
direction corresponding to the stripe orientation at that point. This
predicts a greater number of disparity selective neurons for the
direction of disparity corresponding to the stripe orientation as can
be verified electrophysiologically. Also the limits of depth
perception (for example, Panum's fusional area) should be greater in
the direction along the stripes than across as can be tested
psychophysically.

In conclusion, I argued that the orientation of the ocular dominance
stripes optimizes the length of intra-cortical wiring needed to
process binocular disparity for depth perception. This argument
supports the utility of the wiring economy principle as a powerful tool 
in relating organization of the cortex to its function.

\section*{Methods}

I obtained the disparity maps numerically for a grid of points on the
right retina using the following algorithm. I found points in the
visual field that project onto a given point on the right retina and
calculated the location of their image on the left retina. The
difference between the images' coordinates on the left and the right
retinae is binocular disparity. Multiplying it by the magnification
factor I found disparity in cortical coordinates. Including only the
points in the visual field with cortical disparity magnitude in the
interval $[a;b]$ I calculated the frequencies of disparity directions
averaged over all possible directions gaze.

I implemented the algorithm in MATLAB using the following parameters.
The azimuth of the gaze directions as well as the elevation were
uniformly distributed within $30\deg$ of straight ahead.  The
cyclotorsional misalignment was $10\%$ of the elevation
angle.\cite{Bruno} The point of fixation was at finite but large (2000
times interocular) distance. The expression for the cortical
magnification factor was proportional to $1/(e+e_2)$ where $e$ is
eccentricity and $e2=2\deg$.\cite{Tootell} For
the map shown in Fig.\ref{fig:DCa} I chose $a=0.5b$ and $b=0.13e$ 
corresponding to the diameter of the Panum's fusional area.\cite{Hampton}

\section*{Acknowledgments}

I thank M.R. DeWeese, A.M. Zador, J.D. Pettigrew, R.J. Krauzlis,
J.C. Horton, B.G. Cumming, E.M. Callaway, T.D. Albright and, in particular, 
C.F. Stevens for helpful discussions. This research was 
supported by a Sloan Fellowship in Theoretical Neurobiology.


\end{multicols}

\begin{thebibliography}{99}
\small

\bibitem[Barlow {\em et al.}, 1967]{BBP} Barlow, H.B., Blakemore, C. \& Pettigrew, J.D. (1967). The neural mechanism of binocular depth discrimination. J Physiol (Lond) 193, 327-42.

\bibitem[Bauer, 1995]{Bauer} Bauer, H.U. (1995). Development of oriented ocular dominance bands as a consequence of areal geometry. Neural Comput 7, 36-50.

\bibitem[Bruno \& Van den Berg, 1997]{Bruno} Bruno, P. \& van den Berg, A.V. (1997). Relative orientation of primary positions of the two eyes. Vision Res 37, 935-47.

\bibitem[Blasdel \& Campbell, 1998]{Blasdel} Blasdel, G. \& Campbell, D. (1998). Symmetry of cortical magnification in Old and New World primates. Preprint.

\bibitem[Blasdel \& Fitzpatrick, 1984]{BlaFitz} Blasdel, G.G. \& Fitzpatrick, D. (1984). Physiological organization of layer 4 in macaque striate cortex. J Neurosci 4, 880-95.

\bibitem[Cajal, 1995]{Cajal} Cajal, S.R.y. Histology of the nervous system (pp.116-124) Oxford University Press, New-York.

\bibitem[Cherniak, 1992]{Chern} Cherniak, C. (1992). Local optimization of neuron arbors. Biol Cybern 66, 503-10.

\bibitem[Chklovskii \& Koulakov, 1999]{OD} Chklovskii, D.B. \& Koulakov, A.A. (1999). Ocular dominance patterns in mammalian visual cortex: A wirelength optimization approach. Preprint.

\bibitem[Chklovskii \& Stevens, 1999]{Cat} Chklovskii, D.B. \& Stevens, C.F. (1999). Wiring the brain optimally, preprint.

\bibitem[Cowey, 1979]{Cowey} Cowey, A. (1979). Cortical maps and visual perception: the Grindley Memorial Lecture. Q J Exp Psychol 31, 1-17.

\bibitem[Daniel \& Whitteridge, 1961]{DanWhitt} Daniel, P.M. \& Whitteridge, D. (1961). J Physiol(Lond) 159, 203-221.

\bibitem[Enright, 1980]{Enright} Enright, J.T. (1980). Ocular translation and cyclotorsion due to changes in fixation distance. Vision Res 20, 595-601.

\bibitem[Erwin {\em et al.}, 1995]{Erwin} Erwin, E., Obermayer, K. \& Schulten, K. (1995). Models of orientation and ocular dominance columns in the visual cortex: a critical comparison. Neural Comput 7, 425-68.

\bibitem[Goodhill {\em et al.}, 1997]{GoodOrient} Goodhill, G.J., Bates, K.R. \& Montague, P.R. (1997). Influences on the global structure of cortical maps. Proc R Soc Lond B Biol Sci 264, 649-55.

\bibitem[Goodhill \& Willshaw, 1994]{GoodWill}  Goodhill, G.J. \& Willshaw, D.J. (1994). Neural Computation 6, 615-621.

\bibitem[Hampton, 1983]{Hampton} Hampton, D.R. \& Kertesz, A.E. (1983). The extent of Panum's area and the human cortical magnification factor. Perception 12, 161-5.

\bibitem[Helmholtz, 1962]{Helmholtz} Helmholtz, H.L.v. (1962) Helmholtz's treatise on physiological optics.
Translated from the 3d German ed. Edited by James P.C. Southall, Dover, New-York.

\bibitem[Horton \& Hocking, 1996]{HortonVary}  Horton, J.C. \& Hocking, D.R. (1996). Intrinsic variability of ocular dominance column periodicity in normal macaque monkeys. J Neurosci 16, 7228-39.

\bibitem[Horton \& Hocking, 1998]{Horpriv} Horton, J.C. \& Hocking, D.R., private communication.

\bibitem[Hubel \& Freeman, 1977]{HubFree}  Hubel, D.H. \& Freeman, D.C. (1977). Projection into the visual field of ocular dominance columns in macaque monkey. Brain Res 122, 336-43.

\bibitem[Hubel \& Wiesel, 1968]{HubWies68} Hubel, D.H. \& Wiesel, T.N. (1968). Receptive fields and functional architecture of monkey striate cortex. J Physiol (Lond) 195, 215-43.

\bibitem[Hubel \& Wiesel, 1970]{HW70} Hubel, D.H. \& Wiesel, T.N. (1970). Stereoscopic vision in macaque monkey. Cells sensitive to binocular depth in area 18 of the macaque monkey cortex. Nature 225, 41-2.

\bibitem[Hubel \& Wiesel, 1974]{HubWies74} Hubel, D.H. \& Wiesel, T.N. (1974). Uniformity of monkey striate cortex: a parallel relationship between field size, scatter, and magnification factor. J Comp Neurol 158, 295-305.

\bibitem[Jones {\em et al.}, 1991]{Jones} Jones, D.G., Van Sluyters, R.C. \& Murphy, K.M. (1991). A computational model for the overall pattern of ocular dominance. J Neurosci 11, 3794-808.

\bibitem[Kaas, 1997]{Kaas} Kaas, J.H. (1997). Topographic maps are fundamental to sensory processing.
Brain Res Bull 44, 107-12.

\bibitem[LeVay {\em et al.}, 1985]{LeVay} LeVay, S., Connolly, M., Houde, J. \& Van Essen, D.C.  (1985). The complete pattern of ocular dominance stripes in the striate cortex and visual field of the macaque monkey. J Neurosci 5, 486-501.

\bibitem[Minken \& Van Gisbergen, 1994]{Minken}  Minken, A.W. \& Van Gisbergen, J.A. (1994). A three-dimensional analysis of vergence movements at various levels of elevation. Exp Brain Res 101, 331-45.

\bibitem[Mitchison, 1991]{Mitch1} Mitchison, G. (1991). Neuronal branching patterns and the economy of cortical wiring. Proc R Soc Lond B Biol Sci 245, 151-8.

\bibitem[Mok {\em et al.}, 1992]{Mok} Mok, D., Ro, A., Cadera, W., Crawford, J.D. \& Vilis, T. (1992). Rotation of Listing's plane during vergence. Vision Res 32, 2055-64.

\bibitem[Mountcastle, 1957]{Mou} Mountcastle, V.B. (1957). J Neurophysiol 20, 408-434.

\bibitem[Nakayama, 1983]{Nakayama}  Nakayama, K. in Vergence Eye Movements: Basic and Clinical Aspects (eds. Schor, C. \& Ciuffreda, K.J.) 543-566 Butterworths, London.

\bibitem[Poggio \& Fischer, 1977]{PF} Poggio, G.F. \& Fischer, B. (1977). Binocular interaction and depth sensitivity in striate and prestriate cortex of behaving rhesus monkey. J Neurophysiol 40, 1392-405.

\bibitem[Rosa {\em et al.}, 1992]{Rosa}  Rosa, M.G., Gattass, R., Fiorani, M., Jr. \& Soares, J.G. (1992). Laminar, columnar and topographic aspects of ocular dominance in the primary visual cortex of Cebus monkeys. Exp Brain Res 88, 249-64.

\bibitem[Schwartz, 1980]{Schwartz} Schwartz, E.L. (1980). Computational anatomy and functional architecture of striate cortex: a spatial mapping approach to perceptual coding. Vision Res 20, 645-69.

\bibitem[Swindale, 1996]{SwinRev} Swindale, N.V. (1996). The development of topography in the visual cortex: a review of models. Network: Computation in Neural Systems 7, 161-247.

\bibitem[Tootell {\em et al.}, 1988]{Tootell} Tootell, R.B., Switkes, E., Silverman, M.S. \& Hamilton, S.L. (1988). Functional anatomy of macaque striate cortex. II. Retinotopic organization. J Neurosci 8, 1531-68.

\bibitem[Van Rijn \& Van den Berg, 1993]{Van}  Van Rijn, L.J. \& Van den Berg, A.V. (1993). Binocular eye orientation during fixations: Listing's law extended to include eye vergence. Vision Res 33, 691-708.

\bibitem[Volkmann, 1859]{Volkmann} Volkmann, A.W. (1859). Die Stereoskopischen Erscheinungen... J. Graefes Arch. Ophtal 2, 1-100.

\bibitem[Wiesel \& Hubel, 1974]{WiesHub74} Wiesel, T.N., Hubel, D.H. \& Lam, D.M. (1974). Autoradiographic demonstration of ocular-dominance columns in the monkey striate cortex by means of transneuronal transport. Brain Res 79, 273-9.

\bibitem[Wiskott \& Sejnowski, 1998]{Wisk} Wiskott, L. \& Sejnowski, T. (1998). Constrained optimization for neural map formation: a unifying framework for weight growth and normalization. Neural Comput 10, 671-716.

\bibitem[Young, 1992]{Young} Young, M.P. (1992). Objective analysis of the topological organization of the primate cortical visual system. Nature 358, 152-5.

\end{thebibliography}
\end{document}